\documentclass[12pt]{article}
\usepackage{epsfig}
\def\la{\mathrel{\mathpalette\fun <}}
\def\ga{\mathrel{\mathpalette\fun >}}
\def\fun#1#2{\lower3.6pt\vbox{\baselineskip0pt\lineskip.9pt
        \ialign{$\mathsurround=0pt#1\hfill##\hfil$\crcr#2\crcr\sim\crcr}}}

\begin{document}
\pagestyle{empty}

\topmargin -0.4in
\footskip -0.2in


\begin{center}
\begin{large}
{\bf Joint Efficient Dark-energy Investigation (JEDI):\\
a Candidate Implementation of \\
the NASA-DOE Joint Dark Energy Mission (JDEM)}
\end{large}
\end{center}

\vskip 0.5cm

\noindent
Arlin Crotts (Columbia University)\\
Peter Garnavich (University of Notre Dame)\\
William Priedhorsky, Salman Habib, Katrin Heitmann (Los Alamos National Laboratory)\\
Yun Wang, Eddie Baron, David Branch (University of Oklahoma)\\
Harvey Moseley, Alexander Kutyrev (NASA Goddard Space Flight Center)\\
Chris Blake (University of British Columbia)\\
Edward Cheng (Conceptual Analytics)\\
Ian Dell'Antonio (Brown University)\\
John Mackenty (Space Telescope Institute)\\
Gordon Squires (California Institute of Technology)\\
Max Tegmark (Massachusetts Institute of Technology)\\
Craig Wheeler (University of Texas at Austin)\\
Ned Wright (University of California at Los Angeles)

\vskip 0.5cm

Solving the mystery of the nature of dark energy 
is the most important problem in cosmology today. 
JEDI (Joint Efficient Dark-energy Investigation) is a 
candidate implementation of the NASA-DOE Joint Dark Energy Mission (JDEM). 
JEDI will probe dark energy in three independent ways by measuring
the expansion history of the universe:
(1) using type Ia supernovae as cosmological standard candles
over a range of distances,
(2) using baryon oscillations as a cosmological standard ruler over a
range of cosmic epochs,
(3) mapping the weak gravitational lensing distortion by foreground galaxies
of the images of background galaxies at different distances.
JEDI will unravel the nature of dark energy with accuracy and precision.
 
JEDI is a 2m-class space telescope 
with the unique ability of simultaneous wide-field 
imaging (0.8-4.2$\,\mu$m in five bands) and multi-slit 
spectroscopy (0.8-3.2$\,\mu$m) with a field of view of 1 square degree.
What makes JEDI efficient is its ability to simultaneously obtain high
signal-to-noise ratio, moderate resolution slit spectra for all
supernovae and $\sim$ 5000 galaxies in its wide field of view, and 
to combine imaging and spectroscopy so that the appropriate balance 
of time is devoted to each. 
Another unique feature of JEDI is that it will be cold. This extends the
wavelength range just past 4$\,\mu$m which is critical in limiting systematics
caused by dust extinction.
JEDI requires 64 2048$\times$2048 HgCdTe detectors, and 
8 175$\times$384 arrays of microshutters  
as the spectrograph slit mask (both already developed for the JWST).
JEDI will orbit in L2, with sunshields to achieve passive cooling.

JEDI will measure the cosmic expansion history $H(z)$ as a free 
function to $\la$ 2\% accuracy in redshift bins of 0.2-0.3. 
Assuming a flat universe and $\sigma(\Omega_m)=0.01$ (0.03),
JEDI could measure a departure from a vanilla $\Lambda$CDM
model ($w_0=-1$, $w'=0$) with $\sigma(w_0)=0.013$ (0.031) and 
$\sigma(w')=0.046$ (0.063). 
JEDI will obtain the well-sampled 
lightcurves in $Z, J, H, K, L$ bands and spectra of $\sim$ 14,000 type Ia
supernovae with redshifts ranging from 0 to 1.7;
the redshifts of $\sim$ 10-100 million galaxies to $H\sim 23$ and $z\sim 4$
over 1000-10,000 square degrees; and measurements of the shapes of galaxies 
over 1000-10,000 square degrees in $Z,J,H,K,L$ for $\ga 10^9$ galaxies to 
$H\sim 25$.

\newpage
\pagestyle{plain}
\pagenumbering{arabic}
\topmargin -0.9in
\footskip 0.4in

\section{Overview}

\subsection{Science Goals}

The objective of JEDI is to answer the following question:
what is the nature of the dark energy that dominates the universe today?
The primary science goals of JEDI are chosen to ensure that we
answer this fundamental question with accuracy and precision.

\noindent
{\bf Accuracy: statistically significant measurements}

A phenomenological and model-independent approach to unravel
the mystery of dark energy is to measure the cosmic expansion history
accurately. In order to rule out at least 90\% 
of the parameter space (total matter-energy density
versus cosmic time) of allowed dark energy models, we need to
measure the cosmic expansion history $H(z)$
to 2\% accuracy or better in redshift bins with $\Delta z=0.2-0.3$.
This determines the data required by JEDI.

\noindent
{\bf Precision: resolving the systematics}

To ensure that our dark energy measurements are free of systematics,
JEDI will attack the systematics from two fronts:\\
\noindent
(i) Use three independent methods to probe dark energy:\\
\noindent
a) Hubble diagram of type Ia supernovae (SNe Ia):
using SNe Ia as cosmological standard candles at different distances.\\
\noindent
b) Baryon oscillations from galaxy redshift survey:
using baryon oscillations as a cosmological standard ruler at 
different cosmic epochs.\\
\noindent
c) Weak lensing cross correlation cosmography:
using the weak gravitational lensing distortion by foreground galaxies
of the images of background galaxies at different distances.\\
\noindent
(ii) Require sufficient statistics in each method to 
resolve the systematic uncertainties unique to that method.

JEDI requires that the supernova method and the baryon oscillation
method each yield a 2-5\% or better measurement of $H(z)$
(see Figs.{\ref{fig:Hz_SN}}-{\ref{fig:Hz_BO_Blake}}), 
and that the weak lensing method yields a 10\% or better
measurement of $H(z)$, in $\Delta z=0.2-0.3$ redshift bins. 
The large quantity of data from each method will allow us to
study the systematics of each with robust statistics.
The consistency of the dark energy constraints from the
three independent methods will provide a powerful cross-check
(see Figs.{\ref{fig:wwp_sigOmd03}}-{\ref{fig:wwp_sigOmd01}}).
The combination of JEDI data with other complementary data sets,
for example, cosmic microwave background anisotropy data from Planck 
(which will be available by the time JEDI is launched), will provide
even tighter constraints on dark energy (see Fig.{\ref{fig:wwp_sigOmd01}}).

\subsection{Mission Concept}

JEDI is a 2m-class space telescope capable of 
simultaneous wide-field imaging and multiple object spectroscopy.
The JEDI mission concept is motivated by the desire to maximize the
efficiency in SN Ia spectroscopy, and the need to keep the instrumentation
as simple as possible.

The slit spectrograph is motivated by the need to characterize SNe Ia and
obtain their redshifts from their host galaxies.
The same instrument will deliver millions of additional galaxy redshifts.
The amount of exposure time required for spectroscopy for a
SN Ia at $z=1.7$ is about 25 times the time required for imaging.
To maximize efficiency, we should maximize the field of view
(FOV) of the multiple object
spectrograph, such that the spectra of at least 10 new SNe Ia can be obtained
at each visit to a field. Given detector constraints, we have chosen
a FOV of 1 square degree.

Fig.{\ref{fig:Arlin}} shows a strawman focal plane layout for JEDI.
The JEDI focal plane is centered by an imaging array consisting 
of five strips of NIR imaging detectors, covering the wavelength range of 
0.8-4.2 $\mu$m. On either side of the imaging array are 8 spectrograph fields, 
each consists of a 175x384 array of microshutters (each with a slit size 
2$^{\prime\prime}$x5$^{\prime\prime}$) 
--- twice as many as will be used by JWST.
The wavelength coverage of the spectrograph fields is 0.8-3.2 $\mu$m.
JEDI uses 64 HAWAII-2 2048x2048, and 16 HAWAII-1 (1024x512 used area)
HgCdTe detectors.

This focal plane layout allows simultaneous imaging and spectroscopy. 
Each $Z$, $J$, $H$ imaging detector is 225$^{\prime\prime}$ (36mm) 
on the side, with a plate scale
of 6$^{\prime\prime}.26$/mm. The imaging resolution is 0$^{\prime\prime}$.11/pixel.
Each $K$, $L$ imaging detector is 450$^{\prime\prime}$ (18mm) on the side, with a plate scale
of 25$^{\prime\prime}$/mm. The imaging resolution is 0$^{\prime\prime}$.44/pixel.
The plate scale of the spectrographs is 25$^{\prime\prime}$/mm, with $\sim$ 1
pixel per FWHM sampling. The spectral resolution is
$R = \lambda / \Delta \lambda =300$-1000.

It will take 10 hours for the imager to cover 1 degree in the scanning direction.
Meanwhile, 10 hours is roughly right for getting the spectra
of the z=1.7 SNe Ia. Hence the ratio of the exposure times for imaging 
and spectroscopy is about right by design.
The total exposure time for spectroscopy will be divided into
segments of 2000s to minimize read noise and for dithering.
	
Since the plate scales of the imager (6.26$^{\prime\prime}$/mm) and the spectrographs 
(25$^{\prime\prime}$/mm) differ, the imagers will have to share the focal plane
with pick-off mirrors which reflect the light to the spectrographs and
the $K$, $L$-band imaging detectors.	
The optical design will not be simple, but is tractable at a focal ratio of
around f/25. Fig.{\ref{fig:JEDIoptical}} shows a strawman optical design for
JEDI.

JEDI will orbit in L2, with sunshields to achieve passive cooling.
JEDI will observe 12 fields (1 square degree each in area) every
five days for one year, which yields a sky coverage of 24 square degrees
over two years. This will yield data simultaneously for the SN Ia,
and ultra-deep galaxy redshift, and weak lensing surveys.

JEDI will spend 10 hours per visit on each field.
The spectrographs will obtain the spectra of the hundreds of SNe Ia
(including at least ten new SNe Ia since the previous visit)
in the FOV simultaneously, while the imaging array will image
the entire field by stepping through 1 square degree length of
the field in stepsize equal to the size of one imaging detector.


Very high quality images of SNe Ia will be obtained
within very short exposure times. For example,
a type Ia supernova at z=1.5 reaches at peak brightness
   of J=24.0, and JEDI will obtain a photometric
   signal-to-noise ratio of more than 30 in an hour exposure.
   Fig.{\ref{fig:z1d7_slit2d5as}} shows the simulated 
JEDI spectrum of a SN Ia at $z=1.7$.
This indicate that while we will have very high quality SN Ia
spectra at low and intermediate redshifts, we will have good quality
SN Ia spectra at z=1.7 and beyond (which will be improved
by co-adding the spectra from successive visits).

JEDI will devote one year of observing time on 1000-10,000 square degree
galaxy redshift and weak lensing surveys.
These surveys are carried out simultaneously over the same area, due to
JEDI's unique ability of making simultaneous spectroscopic and imaging
observations.

\subsection{Necessity of Space Observations}

\noindent
{\bf Hubble diagram of type Ia supernovae (SNe Ia)}\\
To derive model-independent constraints on dark energy, it is important
that we precisely measure the cosmic expansion history $H(z)$
in continuous redshift bins from $z\sim 0$ to $z\sim 2$ (the redshift
range in which dark energy is important).
{\it The SNe Ia at $z>1$ are not accessible from the ground,
because the bulk of their light has shifted into the NIR where the sky
background is overwhelming;
hence a space mission is required to probe dark energy using
SNe. Because of JEDI's unique NIR wavelength coverage (0.8-4.2$\mu$m),
we have the additional advantage of observing SNe Ia
in the restframe $J$ band for the entire redshift range
of $0 < z< 1.7$, where they are less affected by dust, and
appear to be better standard candles.}
Since JEDI obtains the spectra of nearly {\it all} supernovae in the
square degree field of view simultaneously, it 
has the capability to obtain the largest number of SNe Ia
at the highest possible redshifts.
SNe Ia occur at higher rates per unit mass in blue spiral galaxies
than in red ones, indicating that at the present epoch a substantial
fraction of SNe Ia are produced by a rather young population, 0.1 to 1
Gyr \cite{Mannucci05}. This fraction should be higher at high
redshift. JEDI will likely observe a significantly larger number
of SNe Ia near and beyond $z=1.7$ than our current conservative
estimate. This will strengthen our ability to detect any evolutionary
effects, and further tighten dark energy constraints.\\

\noindent
{\bf Baryon oscillations from galaxy redshift survey}\\
JEDI uses H$\alpha$ emission line galaxies
to probe baryon oscillation for the entire redshift range
of $0.2 <z < 4$.
This is made possible by our wavelength coverage in the NIR
(0.8 to 3.2$\mu$m for spectroscopy),
which is not possible from the ground.
The H$\alpha$ line is not resonantly scattered and thus is relatively
insensitive to the effects of dust.
Ground-based observations use luminous star-forming galaxies
or luminous elliptical galaxies at $z \sim 1$,
and Lyman break galaxies at $z \sim 3$.
JEDI can measure the expansion history $H(z)$ as 
a free function in $\Delta z=0.2$ redshift slices,
continuously from $z\sim 0.2$ to $z=3.5$.
The JEDI galaxy survey will be less noisy, and less subject to
systematic bias than ground-based surveys. This will enable us to
obtain precise measurements of dark energy density as a free function
of cosmic time.
{\it JEDI has the capability of conducting unpredented redshift surveys
of H$\alpha$ emission line galaxies: a 10,000 (deg)$^2$
to z=2, and a 1000 (deg)$^2$ to z $\sim$ 4 redshift surveys
in one year, as a result of the low background in space 
observations \cite{Glazebrook04} and our powerful multi-slit spectrographs.}
Fig.{\ref{fig:kaos}} shows that a JEDI galaxy redshift survey
improves the constraints on dark energy by factors of a few compared
to a proposed state of the art survey from the ground.
{\it The difference between the baryon oscillation measurements from
JEDI and ground-based surveys is parallel to the difference between
WMAP and pre-WMAP ground-based CMB experiments.}
They are complimentary, and both are important.
The ground-based experiments may provide interesting 
dark energy constraints in the near future, but it will
require a space-based experiment to make precise and definitive measurements
of dark energy density (see Fig.{\ref{fig:Hz_BO_Blake}} and
Fig.{\ref{fig:kaos}}).\\

\noindent
{\bf Weak lensing cross correlation cosmography}\\
The success of a weak lensing survey is primarily determined by how
well the point spread function (PSF) can be modeled, which in turn
depends on the PSF stability. {\it Space observations have the potential
of achieving much more stable and smaller PSF than ground-based observations,
since seeing is not an issue from space.}
A smaller PSF means that the background galaxies are more cleanly 
separated, and that more objects will be resolved (this 
means a larger fraction of objects can be used for gravitational lensing).
Because of JEDI's unique NIR coverage (0.8-4.2$\mu$m for imaging),
we have the added advantage of measuring galaxy shapes in the
restframe visible band (where the intrinsic galaxy shapes are more regular).
A space-based weak lensing survey allows us to achieve accuracy and
precision in constraining dark energy.

\section{Required Precursor Observations, Developments, and Fundamental
Calibrations}

In order to achieve its science goals,
JEDI does not require precursor observations (other than those
required for the fundamental calibration of nearby type Ia supernovae).
The training set of spectroscopic redshifts for photometric redshifts will
be obtained simultaneously as the JEDI photometric observations.

The primary precursor developments required by JEDI are technological
in nature. These are the manufacturing and testing of 64 
2048$\times$2048 and 16 1024$\times$512 HgCdTe detectors, 
and 8 175$\times$384 arrays of microshutters.

The technology for the HgCdTe detectors is mature and proven,
the only issue here is the timely delivery of the relatively
large quantity of detectors, which places requirements on the
funding time table.
The microshutters have been selected by the James Webb Space Telescope (JWST),
and are being developed by Harvey Mosley (PI) et al.
The microshutter technology is progressing
steadily in its Technological Readiness Level (TRL), at TRL 5 of 6
needed for flight, with timely and successful delivery of 4 175$\times$384
microshutter arrays to the JWST expected.

If JEDI were to use only its own supernova data for dark energy analysis
(which was assumed in Fig.{\ref{fig:Hz_SN}}), then
no absolute flux calibration would be needed.  In this case a relative sensitivity
measurement good to 1\% across JEDI's wavelength range would be sufficient.
However, the low-redshift supernovae are important in comparing with the
JEDI events and must be observed in the optical bands.  This means an absolute
calibration and a means of testing the stability of the calibration over the mission
lifetime must be developed. 
We plan to study this issue in detail within the next year, in order
to determine the optimal precursor observations that will allow us
to achieve the required level of fundamental calibration.

Ground-based telescopes are capable of observing in the optical bands
for supernovae at $z<1$ so JEDI does not add the complication
of optical detectors. However, the volume of supernovae coming from JEDI
will require significant time on large ground telescopes for optical follow-up.

\section{Expected error budget}

\noindent
{\bf Supernovae}\\
The raw peak brightness of SN~Ia vary by more than a factor of two,
but using the light curve shape and color information the
dispersion in distance estimation is reduced to $\sim\,$0.16 mag.
Some methods claim dispersions as small as $0.10$ mag are achievable.
If all of this scatter were confidently determined to be caused by a random
process such as the viewing angle of slightly asymmetric explosions,
then reaching the desired error
floor of 2\% would be a trival average of all SNe Ia at a given redshift.
While asymmetry must contribute to the dispersion (given observed intrinsic
polarization in some events), There are a number of other effects which may
be a source of dispersion but vary with cosmic time and may result in systematic
error. These include:\\
\noindent   
{\it Extinction.} Currently extinction by host galaxy dust is estimated by
knowing the intrinsic color of SNe Ia and applying a standard reddening law.
There is some controversy whether the standard Galactic reddening law applies
and a variation in the host dust properties can contribute to the observed dispersion.
Observing over a wide range of wavelengths that include the near infrared, where
dust extinction is less important, allows an independent estimate of the extinction law.
We will observe the rest-frame J band light curves for all the JEDI SNe Ia
by having a cold telescope that reaches 4.2$\,\mu$m. Precursor studies are being
done by the Carnegie supernova group and the ESSENCE collaboration plans
3.6$\,\mu$m Spitzer observations of SNe Ia this fall.  
According to \cite{Cardelli89}, $A_J/A_V=0.282$.
The extinction by the Milky Way can be minimized by choosing
the observing fields to be near the ecliptic poles.
The overall effect of extinction can be reduced
to less than 1\%.\\
\noindent
{\it K-Correction.}
Photometry of the SNe Ia at different redshifts in different color bands 
needs to be mapped onto a consistent rest-frame band.
We will use a set of rest-frame template spectra to build
a rest-frame Spectral Energy Distribution (SED) for the entire
SN Ia wavelength range.
Because of the extraordinary efficiency of the JEDI spectrographs,
we will have at least several high signal-to-noise ratio, moderate resolution
spectra per SN Ia at different epochs.
This will enable us to calibrate and improve our template SED, and
reduce the uncertainty in K-corrections
to around 0.01 mag.
Furthermore, the JEDI bands are spaced in wavelength so that they transform
into each other at certain redshifts, so that uncertainties in K-correction
identically disappear.
The redshift-dependent systematic bias due to K-correction uncertainties
can be reduced to a negligible level by optimizing the analysis
technique; Ref.\cite{WangTegmark05} developed a method that effectively
reduces the global systematic bias (over the entire redshift range)
to a local bias (in each redshift bin) with a much smaller amplitude. \\
\noindent  
{\it Weak lensing.}
The bending of the light from SNe Ia by intervening matter will
modify the observed brightness of SNe Ia.
It has been demonstrated that the effect of weak lensing of SNe Ia 
can be reduced to an negligible level given sufficiently good statistics
\cite{Wang00,WangMukherjee,WangTegmark05}.
For JEDI, this will be smaller than 0.2\%.\\
\noindent  
{\it Selection bias.}
Krisciunas et al. (2005) has shown the importance of searching at a redshift
limit, not a magnitude limit. To avoid Malmquist-like biases, it is necessary to
search a magnitude deeper than the typical peak supernova brightness to sample
the fast-decaying events. JEDI exposure times are designed to detect fainter-than-typical
events at $z=1.7$.\\
\noindent
{\it Gray dust.}
Models of intergalactic gray dust show that it can not be truly gray and must have some
reddening effect. JEDI's broad wavelength coverage (0.8-4.2 $\mu$m) will allow us to tightly
constrain the possibility of gray dust.\\
\noindent
{\it Supernova peak luminosity evolution.}
Models suggest that progenitor metallicity or mass may influence
the peak luminosity of SN~Ia at a fixed light curve decline rate, although, the
models do not agree on the amplitude or even the sign of these effects. Since
both progenitor population metallicity and age vary over cosmic time this is
an important and uncertain source of systematic error.
Progenitor metallicity may be correlated with ultraviolet flux before maximum,
but no observations have demonstrated this effect. There is a strong correlation
between light curve decline rate and host morphology (and host star-formation
rate), but Gallagher et al. (2005) have shown that there is no strong residual correlation
between host  metallicity or star-formation history after correction for light curve shape
and dust extinction. With the current uncertain state of theory and observation, our
plan is to obtain a sufficient range and details in our observations to allow a study
of environmental influences at all redshifts. 
With the over 14,000 SNe Ia (with well sampled light curves and good quality
spectra) from JEDI, we will be able to subtype the SNe Ia, and search for
most possible evolutionary effects.\\
\noindent  
{\it Other possible systematics.}
Because JEDI will obtain over 14,000 SNe Ia, all with well-sampled
lightcurves and good-quality spectra, in an ultra-deep
survey of 24 square degrees, we will not be contaminated by 
non type Ia supernovae. For example, we will be able to differentiate
SNe Ia from Ibc.
The spectra will not only be used for typing and
redshift measurement. The quality of the slit spectra will allow spectral features
and velocities to be correlated with light curve properties so that the distance
scatter can be reduced to an acceptable level. \\

\noindent
{\bf Baryon Oscillations}\\
The systematics of the baryon oscillation method are not
well-understood at present.
Theoretical studies indicate that this method is likely to be substantially
free of systematics \cite{Blake03,Eisenstein04,Glazebrook05}. However, 
systematic effects will inevitably be present in real data; they can only 
be addressed by means of studying mock catalogues constructed from 
realistic cosmological volume simulations \cite{Springel05}.
We plan to quantify the error budget for this method
within the next year or so.
An alternative to the detailed modeling of realistic systematics
(such as those from the inaccurate modeling of redshift distortions)
is to use robust and conservative statistical methods that are
largely free of systematics. For example, Ref.\cite{Blake03}
uses Monte Carlo methods, and divides out the overall
shape of $P(k)$ to reduce sensitivity to systematic
uncertainties (only the oscillations are fitted).
All JEDI baryon oscillation constraints shown in this paper
result from this conservative approach.
Ultimately, we will model the systematics in detail, in order
to derive the most stringent constraints on dark energy.
This would typically improve the accuracy of estimated parameters
by 30-50\% \cite{Glazebrook05}.\\

\noindent
{\bf Weak Lensing}\\
The understanding of the weak lensing systematics is still in its early
stages \cite{Huterer05}. The sources of systematic errors include 
multiplicative and additive errors in the measurements of shear,
and photometric redshift errors.
In order for the systematics not to dominate the cosmological parameter
error budget, Ref.\cite{Huterer05} has shown that the multiplicative
error in shear needs to be smaller than 1\%$\,(f_{sky}/0.025)^{-1/2}$
of the mean shear in any given redshift bin, while the centroids
of photometric redshift bins need to be known to better than 
0.003 $\,(f_{sky}/0.025)^{-1/2}$. 
JEDI can achieve these requirements. Note that since we will
have over 10 million spectroscopic redshifts over the 1000 square
degrees in which weak lensing measurements are made, we
should easily obtain photometric redshifts which surpass
the required accuracy. This is important, since self-calibration (deriving the
systematic biases self-consistently from the data) leads to
a factor of two degradation in cosmological parameter errors.\cite{Huterer05}

\section{How JEDI will quantify our understanding of dark energy}

JEDI will measure the cosmic expansion history $H(z)$ as a free function using
three independent and complementary methods.

\begin{figure}[!hbt]
\begin{center} 
\psfig{file=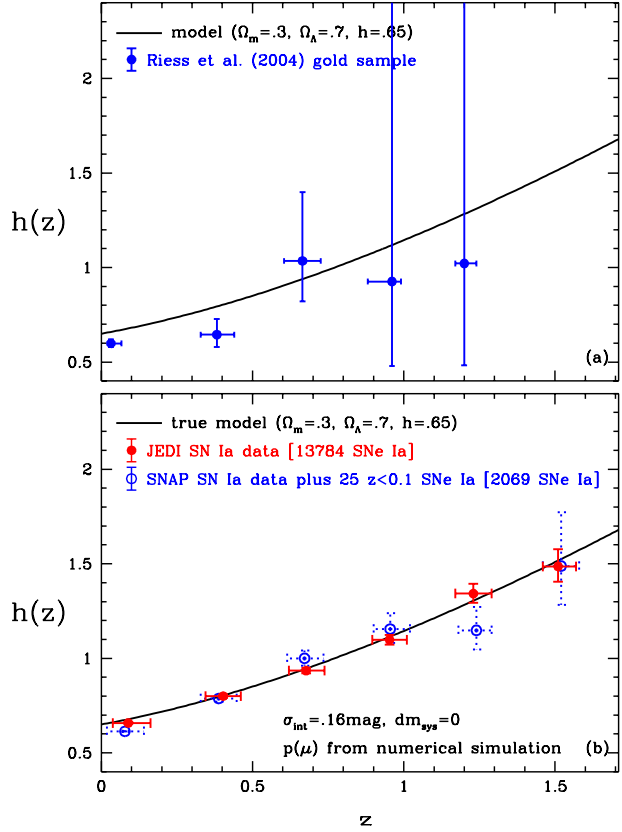,width=11cm}
\vskip -0.3cm
\caption{\footnotesize%
\label{fig:Hz_SN}
The cosmic expansion history measured in uncorrelated redshift
bins from (a) current data, and (b) simulated JEDI supernova data.
Note that $h(z)=H(z)/[100\,$km$\,$s$^{-1}$Mpc$^{-1}]$.
Known systematic uncertainties (intrinsic dispersion in supernova 
peak brightness and weak lensing due to cosmic large scale structure)
are included. We have assumed a flat universe and
that $\Omega_m$ is known.
(Wang \& Tegmark 2005) }
\end{center} 
\end{figure}

JEDI will obtain over 14,000 SNe Ia with well sampled light curves
and good quality spectra. This will allow us to examine
SN Ia systematic uncertainties in great detail, improve the calibration
of SNe Ia as cosmological standard candles, and measure
the cosmic expansion history $H(z)$ to 5\% accuracy or better
($\Delta z=0.28$, see Fig.1);
this gives an accurate measurement of 
the dark energy density as a free function of cosmic time
if $\Omega_m$ is accurately measured from other data (for example,
CMB data from Planck, or JEDI galaxy redshift data).
Fig.{\ref{fig:Hz_SN}} shows uncorrelated estimates of $H(z)$
expected from JEDI SN Ia data, assuming a flat universe, and including known systematics
of SNe Ia as cosmological standard candles (intrinsic scatter in peak brightness
of 0.16 mag and realistic weak lensing by cosmic large scale structure).
The number of SNe Ia assumed correspond to the most conservative estimate
of the SN Ia rate (for a SN Ia delay time of 3.5 Gyr).
We have assumed that $\Omega_m$ is known. Assuming a realistic
prior on $\Omega_m$ (say, a Gaussian with a width of 0.03) will introduce 
some correlation in the $H(z)$ measured in different redshift bins, but 
will not change the size of the error bars significantly.

\begin{figure}[!hbt]
\begin{center} 
\psfig{file=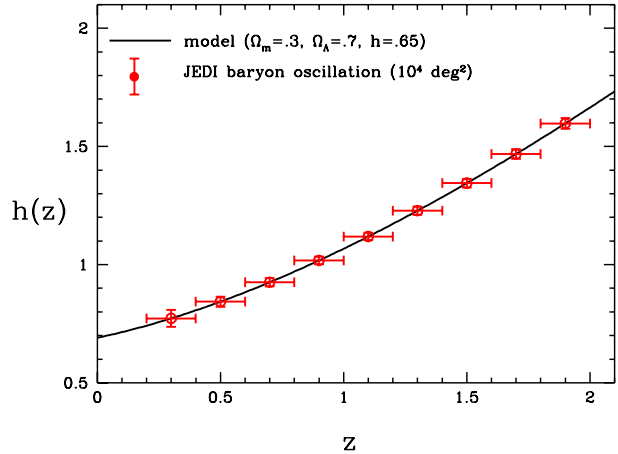,width=11cm}
\vskip -4cm
\caption{\footnotesize%
\label{fig:Hz_BO_Blake}
The estimated 1-$\sigma$ uncertainty on the 
cosmic expansion history measured in uncorrelated redshift
bins from JEDI baryon oscillation data.
The errors are derived using a Monte Carlo method that
largely avoid systematic uncertainties (Blake \& Glazebrook 2003),
and are not sensitive to values of $\Omega_m$ and $\Omega_b$ assumed. 
We have assumed the sound horizon at last scattering as
measured by WMAP \cite{Spergel03}.
See Table 1 for percentage errors in $H^{-1}(z)/s$.}
\end{center} 
\end{figure}

\begin{figure}[!hbt]
\begin{center} 
\psfig{file=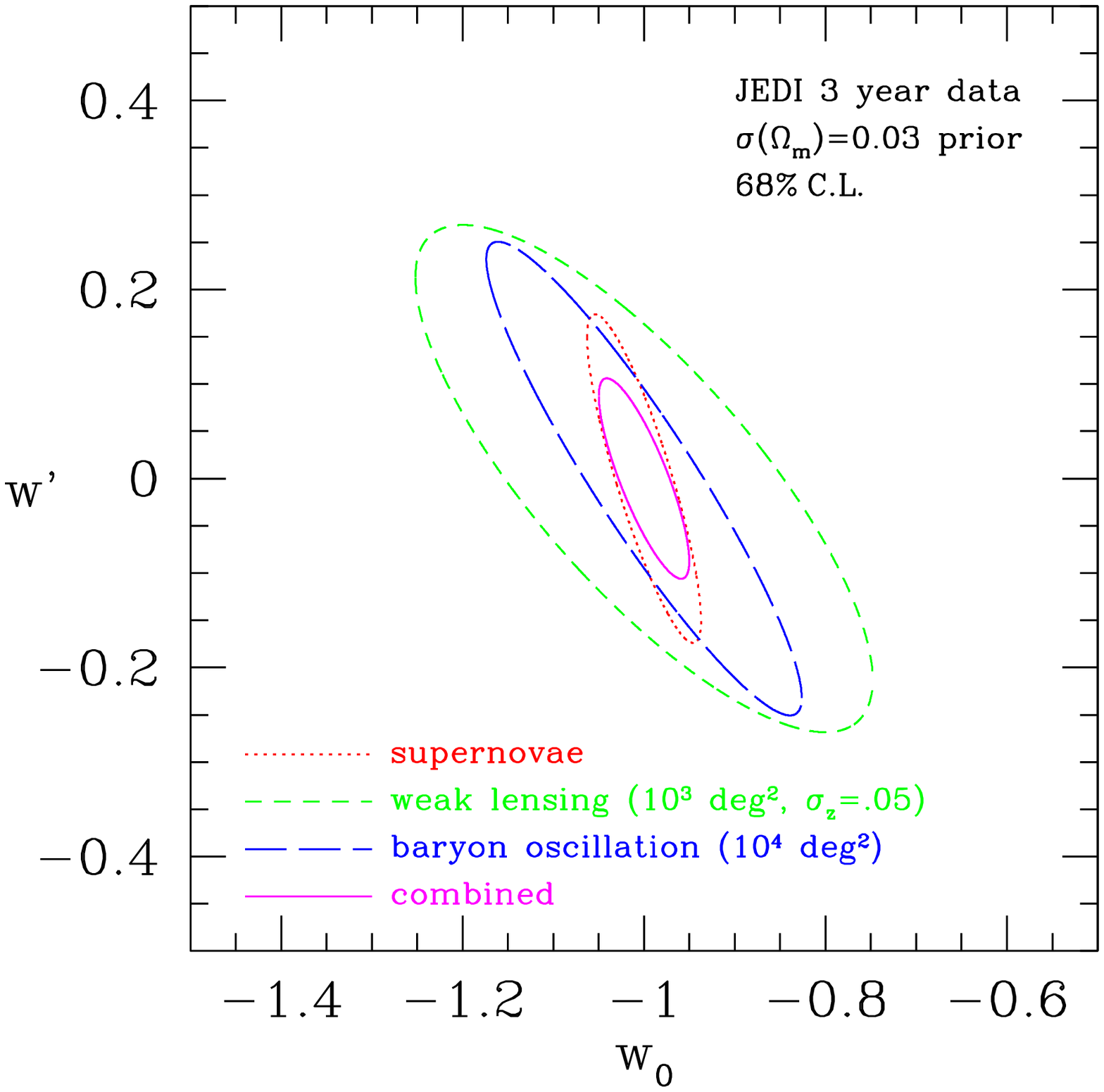,width=10cm}
\vskip -0.3cm
\caption{\footnotesize%
\label{fig:wwp_sigOmd03}
Estimated JEDI measurements on departures from a $\Lambda$CDM model with $w_0=-1$
and $w'=0$, for a prior of $\sigma(\Omega_m)=0.03$ (close to WMAP constraints). 
The baryon oscillation constraint also assumes that
$\Omega_m h^2$ is measured to 3\% accuracy.
The contours indicate the 68.3\% confidence level. See also Table 2.}
\end{center} 
\end{figure}

\begin{figure}[!hbt]
\begin{center} 
\psfig{file=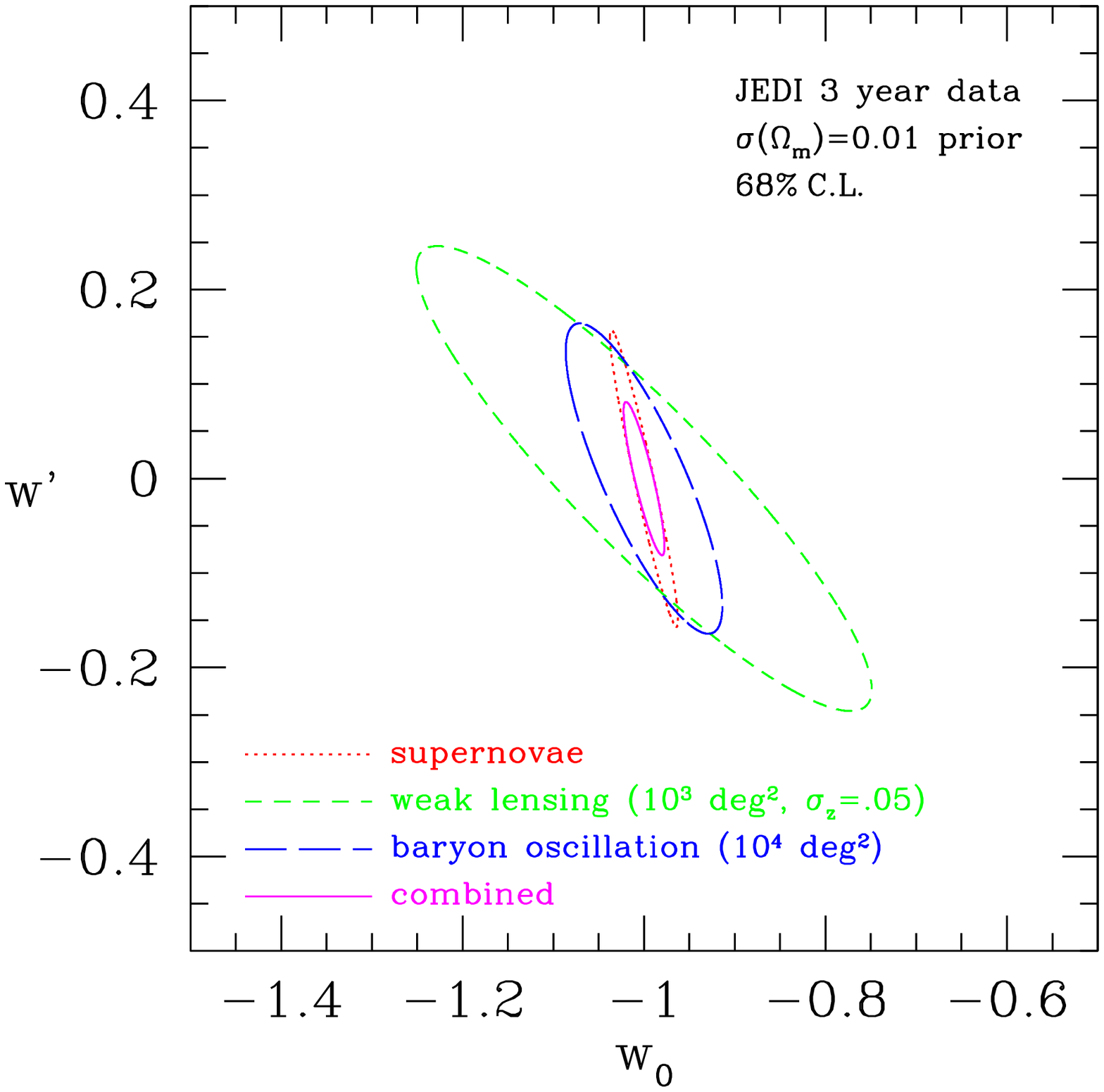,width=10cm}
\vskip -0.3cm
\caption{\footnotesize%
\label{fig:wwp_sigOmd01}
Estimated JEDI measurements on departures from a $\Lambda$CDM model with $w_0=-1$
and $w'=0$, for a prior of $\sigma(\Omega_m)=0.01$ (close to Planck constraints).
The baryon oscillation constraint also assumes that
$\Omega_m h^2$ is measured to 3\% accuracy. 
The contours indicate the 68.3\% confidence level. See also Table 2.}
\end{center} 
\end{figure}

\begin{figure}[!hbt]
\begin{center} 
\psfig{file=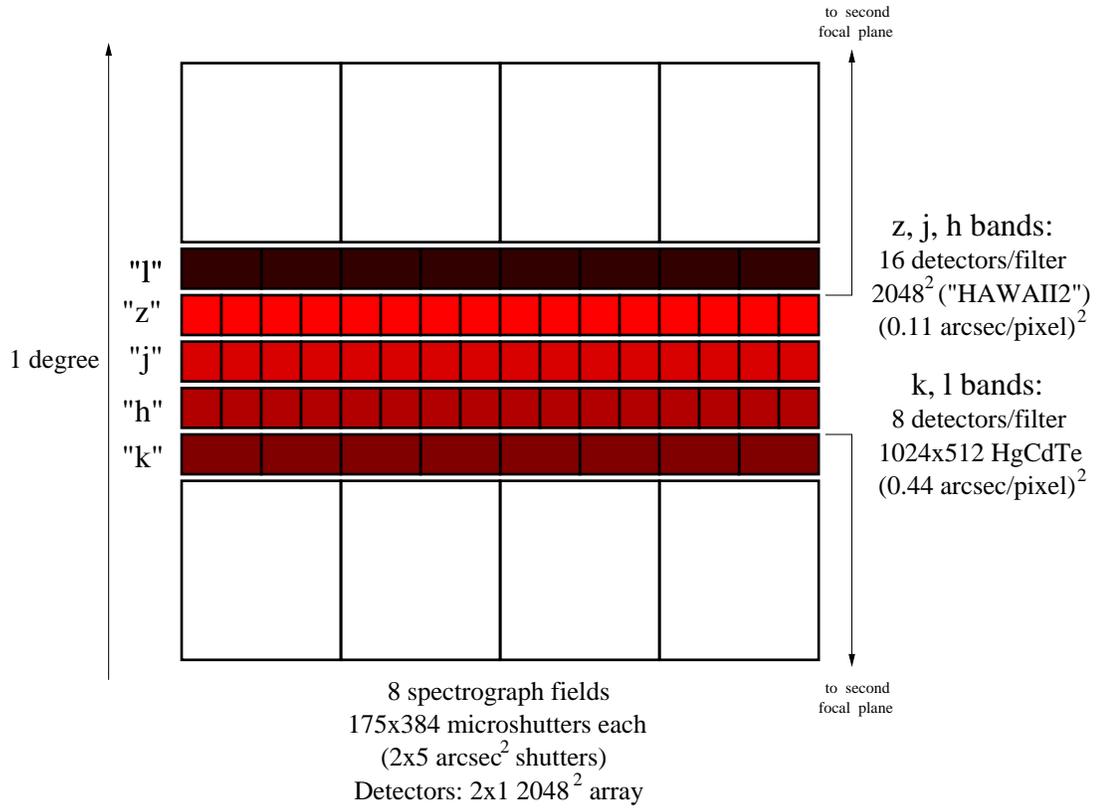,width=12cm,angle=270}
\vskip -0.6cm
\caption{\footnotesize%
\label{fig:Arlin}
JEDI strawman focal plane design.
} 
\end{center} 
\end{figure}

\begin{figure}[!hbt]
\begin{center} 
\psfig{file=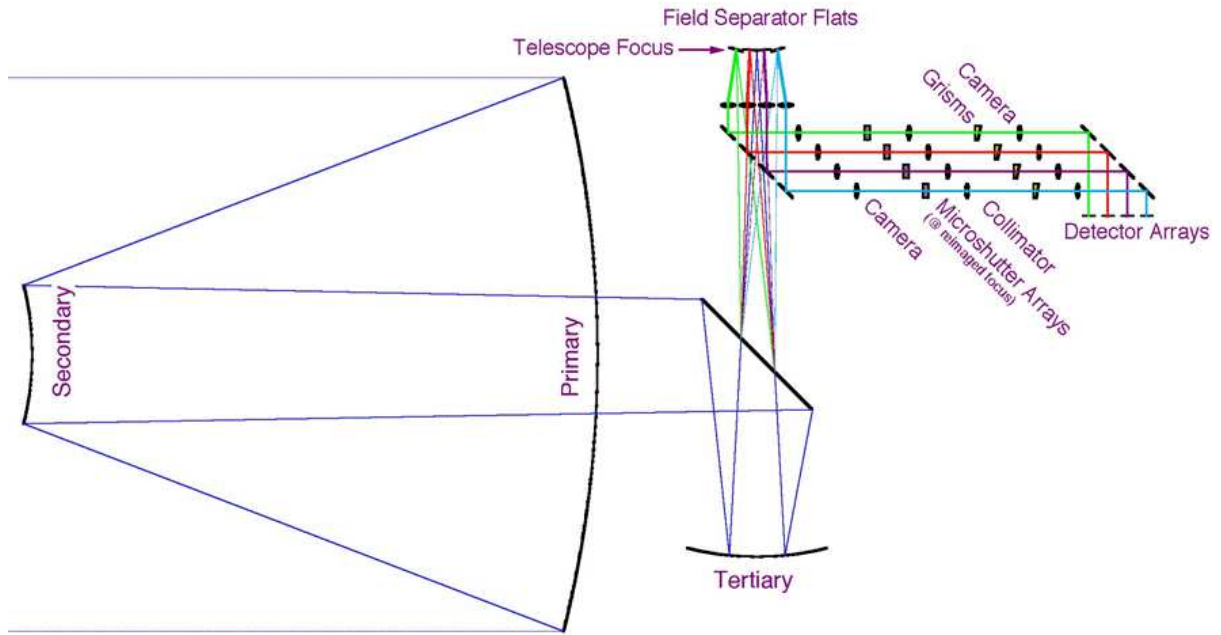,width=16cm}
\caption{\footnotesize%
\label{fig:JEDIoptical}
JEDI strawman optical design (courtesy of Dominic Benford).
For simplicity we have suppressed the K and L band detectors which share the
second focal plane with the microshutter arrays.
} 
\end{center} 
\end{figure}

\begin{figure}[!hbt]
\begin{center} 
\psfig{file=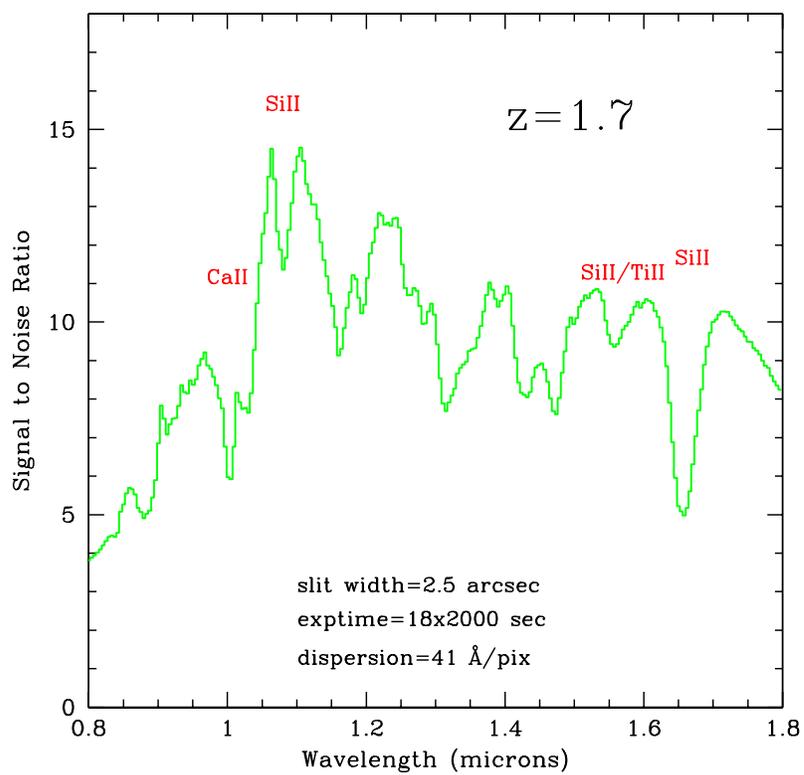,width=11cm}\\
\psfig{file=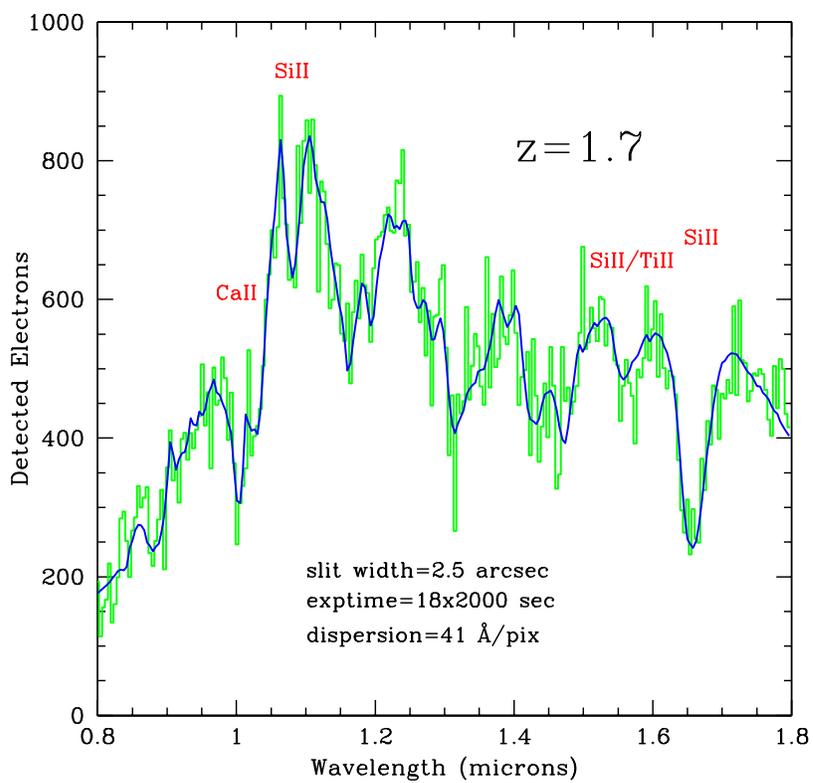,width=11cm}\\
\vskip -0.6cm
\caption{\footnotesize%
\label{fig:z1d7_slit2d5as}
Simulated JEDI spectrum of a SN Ia at $z=1.7$.
} 
\end{center} 
\end{figure}

\begin{figure}[!hbt]
\begin{center} 
\psfig{file=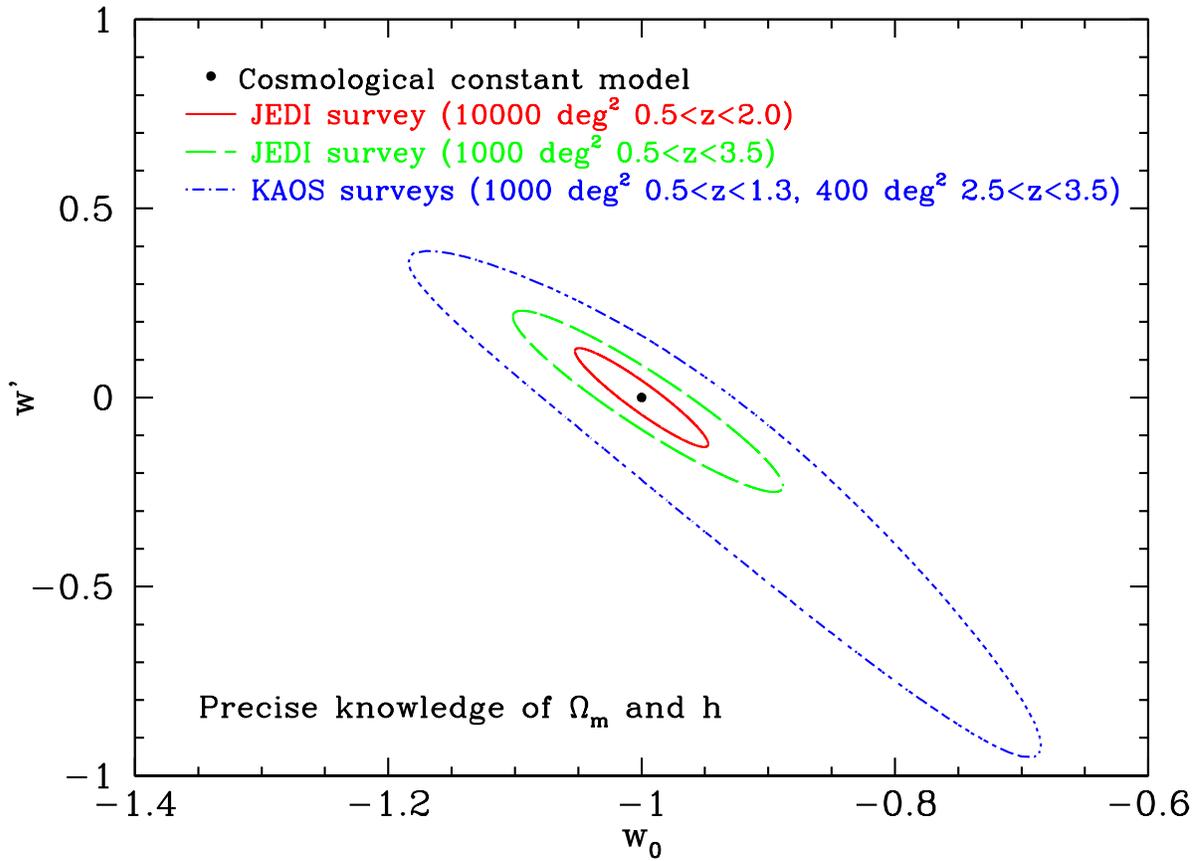,width=12cm,angle=270}
\vskip -0.3cm
\caption{\footnotesize%
\label{fig:kaos}
Comparison of the dark energy constraints derived from JEDI baryon oscillation
measurements and a proposed ground-based survey.}
\end{center} 
\end{figure}

JEDI will obtain around 100 million galaxy redshifts over 10,000 (deg)$^2$
to $z =2$. This will allow us to measure $H(z)$ to 2-3\% accuracy with
$\Delta z=0.2$ (see Table 1).
Fig.{\ref{fig:Hz_BO_Blake}} shows the estimated 1-$\sigma$ uncertainty 
on $H(z)$ measured in uncorrelated redshift
bins from JEDI baryon oscillation data, using 
the conservative Monte Carlo method described in Ref.{\cite{Blake03}}
that is largely free of systematics.
Reasonable modeling of systematics will reduce the error bars by 
30-50\% \cite{Glazebrook05}.
We have assumed the sound horizon at last scattering as
measured by WMAP \cite{Spergel03}, and that
the comoving number density of observed galaxies 
is such that $nP=3$ \cite{Blake03}.
A byproduct of the 1000 (deg)$^2$ weak lensing survey
is a deeper galaxy redshift survey, at least 10 million galaxy redshifts 
over 1000 (deg)$^2$ to $z\sim 4$. This will extend our measurement of $H(z)$
to $z \sim 4$ at 3-4\% accuracy.
Note that Fig.{\ref{fig:Hz_SN}} and Fig.{\ref{fig:Hz_BO_Blake}}
have errors in $H(z)$ with opposite trends in $z$, yielding a combined
accuracy of $\la$ 2\% in $H(z)$.

JEDI will obtain images of galaxies over 1000 (deg)$^2$ to a magnitude
of $H\sim 25$. A shallower weak lensing survey over 10,000 (deg)$^2$
will be a byproduct of the 10,000 (deg)$^2$ galaxy redshift survey for 
baryon oscillations.
These can be used for weak lensing measurements.
It has been shown that a space-based weak lensing survey generally
provides less stringent constraints on dark energy than a space-based
supernova survey \cite{Zhang03}. However, it provides a powerful cross-check
of the other two methods, which is important in arriving at accurate
measurements of dark energy density free of systematic biases.
We plan to conduct detailed studies within the next year
to derive the error on $H(z)$ expected from the
JEDI weak lensing survey.

In order to compare with other experiments, 
Figs.{\ref{fig:wwp_sigOmd03}}-{\ref{fig:wwp_sigOmd01}} show
the estimated JEDI measurements on departures from a $\Lambda$CDM model 
with $w_0=-1$ and $w'=0$ (see also Table 2).
The weak lensing constraints have been estimated by scaling the results
of Ref.\cite{Zhang03}.

\section{Project's risk areas and strengths}

\noindent
{\it Risk Areas.}
JEDI's risk area is in the requirement of 64 2048$\times$2048 HgCdTe detectors, and 
8 175$\times$384 arrays of microshutters  
as the spectrograph slit mask.
Both the 2048$\times$2048 HgCdTe detectors and the 175$\times$384 microshutter arrays 
are already developed for the JWST.
The detector technology is mature; the risk arises from the relatively large number
of detectors that JEDI requires. Given sufficient funding and time, the risk in
the detectors being successfully delivered will diminish.
We have not yet packaged all of the instrumentation in the spacecraft yet, but
estimate that this is possible given existing spacecraft bus and shroud
architecture available within the likely mission cost caps.
The microshutter arrays currently are progressing along the technological readiness
level on schedule for expected timely delivery to the JWST. To avoid unforeseeable
risks in this new technology, we will investigate alternatives as backup in order to
minimize risk.
Our industrial partner, Lockheed Martin Corporation, is helping
us examine these issues. \\
\noindent
{\it Strengths.}
JEDI's strengths are in its extraordinary efficiency that allows
an ambitious scientific program to be carried out.
JEDI has the unique ability of simultaneously obtaining slit 
spectra for all objects in the wide field of view. Slit spectra will
provide a higher signal-to-noise ratio at a better resolution than
grism spectroscopy.
It will obtain over 14,000 SNe Ia with well sampled light curves
and good quality spectra in two years,
and 10-100 million galaxy spectra over 1000-10,000 square degrees
(with imaging in the same area) in one year.
It will measure dark energy density as a free function using three powerful
and independent methods:
(1) using type Ia supernovae as cosmological standard candles
at different distances,
(2) using baryon oscillations as a cosmological standard ruler at 
different cosmic epochs,
(3) using the weak gravitational lensing distortion by foreground galaxies
of the images of background galaxies at different distances.
JEDI will unravel the nature of dark energy with accuracy and precision
(see Figs.{\ref{fig:Hz_SN}}-{\ref{fig:wwp_sigOmd01}} and
Tables 1-2).
JEDI will have the unique ability to triangulate the dark energy properties,
and avoid pitfalls arising from systematics of any one or two methods
(see Figs.{\ref{fig:wwp_sigOmd03}}-{\ref{fig:wwp_sigOmd01}}).
JEDI is the robust project in terms of changes in our understanding of
cosmological probes, and fills a generally useful niche by exploiting the
lowest-noise region in terms of background at 1-4 microns, where
moderate-redshift galaxy and SN science is most compelling.

\section{Other Issues}


\noindent
{\bf Technology R\&D requirements}\\
As noted above, the microshutter and NIR detector technology needed
     for JEDI is being developed for the JWST.
However, R\&D will help minimize risk.\\
\noindent
{\bf Relation to JDEM}\\ 
JEDI is a candidate mission concept for JDEM.\\
\noindent
{\bf Access to facilities and other instrumentation}\\
Since JEDI focuses on the near IR (0.8 to 4.2$\,\mu$m) which is optimal for
high-redshift supernovae, we will need
to coordinate with ground based optical surveys. These include
the LSST, PanSTARRS, ALPACA, DES, etc.\\
\noindent
{\bf Timeline for the completion of the experiment}\\
The core science of JEDI will require a mission duration of 3-5 years, with a
launch as early as 2012.

\vspace{3in}
\noindent
\footnotesize{* We thank Pier-Stefano Corasaniti
for sending us the Fisher matrices of $(w_0, w')$ for the JEDI SNe Ia.}

\newpage

\vspace{0.5in}
\noindent
Table 1: The percentage errors in the Hubble parameter $H(z)$
and the comoving distance $r(z)$ expected for the JEDI 10,000 (deg)$^2$
to $z=2$ redshift survey. We have assumed a comoving number
density of galaxies such that $nP=3$ (where $P$ is the power spectrum);
$n b^2$ is the corresponding surface density of galaxies (per square arcmin). 
Note that $b$ is the linear bias factor, and $s$ is the sound horizon
at decoupling (measured to 1.36\% by WMAP \cite{Spergel03}).

\vspace{0.2in}
\begin{tabular}{|l|lll|}
\hline
redshift  &  $n b^2$  & $ r(z)/s$  & $[c/H(z)]/s$\\
\hline
0.2-0.4 &   0.1  &   4.1 &  6.1 \\
0.4-0.6 &   0.4   &  2.0  & 3.0\\
0.6-0.8 &   0.7  &   1.3  & 1.9\\
0.8-1.0 &   1.1  &   0.9  & 1.3\\
1.0-1.2 &   1.6  &   0.7  & 1.0\\
1.2-1.4 &   2.2   &  0.6  & 1.1\\
1.4-1.6 &   2.7   &  0.6  & 0.9\\
1.6-1.8 &   3.3   &  0.5  & 0.9\\
1.8-2.0 &   3.9   &  0.5  & 0.9\\
\hline
\end{tabular}

\vspace{0.5in}
\noindent
Table 2: JEDI constraints on a departure from a vanilla $\Lambda$CDM
model with $w_0=-1$, $w'=0$. Marginalized errors are shown for
$w_0$ and $w'$ for priors of $\sigma(\Omega_m)$=0.03 and 0.01.
The data used are the supernovae,
weak lensing (1000 square degrees and $\sigma_z=0.05$), and baryon oscillations
(10,000 square degrees to $z=2$).

\vspace{0.2in}
\begin{tabular}{|c|l|l|}
\hline
prior: $\sigma(\Omega_m)$ & $\sigma(w_0)$ & $\sigma(w')$ \\
\hline
 0.03 & 0.031 & 0.063 \\
  		       0.01 & 0.013 & 0.046 \\
  		      \hline

\hline
\end{tabular}

\newpage

\end{document}